\documentclass[lettersize,journal]{IEEEtran}
\usepackage{amsmath,amsfonts}
\usepackage{algorithmic}
\usepackage{array}
\usepackage[caption=false,font=normalsize,labelfont=sf,textfont=sf]{subfig}
\usepackage{textcomp}
\usepackage{stfloats}
\usepackage{url}
\usepackage{verbatim}
\usepackage{graphicx}
\usepackage{threeparttable}
\usepackage{xcolor}
\usepackage{hyperref}
\usepackage{pdfpages}
\ifCLASSINFOpdf
\else
\fi
\begin{document}
\title{Shared-PIM: Enabling Concurrent Computation \\and Data Flow for Faster Processing-in-DRAM}

\author{Ahmed~Mamdouh$^*$,~\IEEEmembership{}
        Haoran~Geng$^*$,~\IEEEmembership{Student Member,~IEEE,}
        Michael~Niemier,~\IEEEmembership{Senior Member,~IEEE,}
        Xiaobo~Sharon~Hu,~\IEEEmembership{Fellow,~IEEE,}
        and~Dayane~Reis,~\IEEEmembership{Senior Member,~IEEE}% <-this % stops a space
\thanks{A. Mamdouh and D. Reis are with the Department
of Computer Science and Engineering, University of South Florida, Tampa,
FL, 33620 USA (e-mail: dayane3@usf.edu). }% <-this % stops a space
\thanks{H. Geng, M. Niemier, and X.S. Hu are with the Department of Computer Science and Engineering, University of Notre Dame. Notre Dame, IN, 46556 USA (e-mail: shu@nd.edu).}% <-this % stops a space
\thanks{$^*$ The authors contributed equally to this work.}
\thanks{Manuscript received August 26, 2024; revised December 13, 2024.}}


\markboth{Journal of \LaTeX\ Class Files,~Vol.~14, No.~8, August~2015}%
{Shell \MakeLowercase{\textit{et al.}}: Bare Demo of IEEEtran.cls for IEEE Journals}



% \includepdf[pages=-]{TCAD_Revision_Letter_Shared_PIM.pdf}

\maketitle

\begin{abstract}
Processing-in-Memory (PIM) enhances memory with computational capabilities, potentially solving energy and latency issues associated with data transfer between memory and processors. However, managing concurrent computation and data flow within the PIM architecture incurs significant latency and energy penalty for applications. This paper introduces Shared-PIM, an architecture for in-DRAM PIM that strategically allocates rows in memory banks, bolstered by memory peripherals, for concurrent processing and data movement. Shared-PIM enables simultaneous computation and data transfer within a memory bank. When compared to LISA, a state-of-the-art architecture that facilitates data transfers for in-DRAM PIM, Shared-PIM reduces data movement latency and energy by 5$\times$ and 1.2$\times$, respectively. Furthermore, when integrated to a state-of-the-art (SOTA) in-DRAM PIM architecture (pLUTo), Shared-PIM achieves 1.4$\times$ faster addition and multiplication, and thereby improves the performance of matrix multiplication (MM) tasks by 40\%, polynomial multiplication (PMM) by 44\%, and numeric number transfer (NTT) tasks by 31\%. Moreover, for graph processing tasks like Breadth-First Search (BFS) and Depth-First Search (DFS), Shared-PIM achieves a 29\% improvement in speed, all with an area overhead of just 7.16\% compared to the baseline pLUTo.
\end{abstract}


% \fi
\IEEEpeerreviewmaketitle

\section{Introduction}

\IEEEPARstart{M}{emory} intensive applications, (e.g., machine learning, data analytics, etc.), require substantial memory resources to facilitate computation~\cite{memory_intensive_app}. A major hurdle for said applications is costly data transfer between a system's memory and processing units \cite{WheretoProcess}. The separation of data storage and computation into distinct physical components results in inherent challenges related to memory bandwidth constraints. This has motivated research in data-centric computing approaches like processing-in-memory (PIM) (e.g., \cite{pluto,PPIMCE,imcrypto,X-poly,asifuzzaman2023survey, seshadri2017ambit, li2017drisa}) where a system's memory is augmented with compute capabilities. Different PIM approaches for dynamic random-access memory (DRAM) have been considered, e.g., pLUTO~\cite{pluto}, AMBIT~\cite{seshadri2017ambit}, DRISA~\cite{li2017drisa}, etc. While each architectural solution has its nuances, they all make direct modifications to the memory cells and periphery circuits in order to allow both logic and arithmetic computations to occur seamlessly within the memory itself. 

% While in-DRAM PIM solutions can improve computational efficiency by performing computation \textit{in-situ}, operations supported in some designs (e.g., in AMBIT and DRISA) are limited to bitwise Boolean logic (e.g., AND, OR, INV, etc.), which requires a multi-step approach for performing more sophisticated computations. While supporting bitwise operations increases the general applicability of a PIM architecture for a wider range of applications, it may also lead to increased latency for operations such as multi-bit additions and multiplications. To combat this challenge, pLUTo \cite{pluto}, a state-of-the-art (SOTA) in-DRAM PIM architecture, implements operations via look-up tables (LUTs) that are stored inside the memory. Nevertheless, pLUTo's reliance on LUT-based computations still poses challenges, particularly for higher-precision operations like 32-bit addition and multiplication. Indeed, due to LUT size requirements, a single DRAM array cannot accommodate large LUTs, thereby necessitating the use of multiple arrays. The multi-array approach exacerbates inter-array data transfer, and diminishes performance benefits derived from PIM solutions. 

%\textcolor{red}{
All existing in-DRAM PIM architectures must process computations and perform data movement operations sequentially within the same array owing to the dual role of the memory as both a storage and a processing unit. This constraint results in significant latency overhead and prevents the simultaneous execution of data processing and transfers. Current in-DRAM PIM fabrics employ data movement frameworks like LISA \cite{lisa} (for fast inter-subarray data movement) and Rowclone \cite{seshadri2013rowclone} (for fast intra-subarray data movement). Despite their advantages, these frameworks are still limited by their inability to support concurrent data movement and computation. This sequential processing increases latency and restricts the full potential of in-DRAM computing.%}% works performance enhancements.}

To help address limitations associated with existing PIM solutions, particularly inter-subarray data movement, this paper introduces Shared-PIM, an architecture for in-DRAM PIM that supports concurrent processing and data transfers. Shared-PIM allocates a few rows in each array in a bank as {\it shared rows} that collaboratively manage data movement and allow for concurrent operation within PIM arrays. To support Shared-PIM operations, we incorporate additional sense amplifiers and an internal bus within the memory bank. In conjunction with the shared rows, these peripheral circuits enable a new memory structure that we refer to as a {\it shared array}. After a computation is initiated within a PIM unit, the shared array is activated, and facilitates simultaneous data movement. Shared-PIM also enables pipelining within the PIM architecture, achieving more efficient computation compared to existing solutions \cite{seshadri2013rowclone, lisa}. The specific contributions of this paper are summarized below:

\begin{enumerate}
    \item We propose Shared-PIM, an architecture for in-DRAM PIM that enables concurrent computation and data movement within a bank through the addition of an internal bus and sense amplifiers. %Shared-PIM facilitates simultaneous computation and data movement, addressing key challenges in existing in-DRAM PIM designs.
    
    \item %\textcolor{blue}{
    We present circuit-level changes to the DRAM cell and peripherals, introducing what we call a \textit{Shared Row}, in conjunction with bank level sense amplifiers to realize the Shared-PIM architecture. %}
    
    \item We have integrated Shared-PIM into the state-of-the-art pLUTo in-DRAM PIM design \cite{pluto} to facilitate inter-array data transfers. This implementation shows that Shared-PIM can be easily leveraged to enhance the latency of existing in-DRAM PIM solutions, with secondary energy benefits.
    
    %\item We have developed a specialized compiler leveraging NVBit \cite{NVBit} to evaluate Shared-PIM for convolutional neural network (CNN), fast Fourier transform (FFT), and breadth first search (BFS) compute tasks. The compiler is designed to optimize parallel operations, ensuring optimal performance and efficient data flow.
\end{enumerate}

%This compiler promotes parallel operations with PIM, while monitoring data dependencies, which guarantees peak performance and streamlined data flow within the system.

We implement and evaluate Shared-PIM at the circuit level through SPICE simulations using ASU's 45nm low-power Predictive Technology Model (PTM) \cite{zhao2007predictive}, and adhering to the JEDEC DDR3-1600 timing standards \cite{jesd792012ddr3}. We compare the inter-subarray copy latency and energy efficiency of Shared-PIM against standard memcpy, Rowclone \cite{seshadri2013rowclone}, and LISA \cite{lisa} integrated with the in-DRAM PIM pLUTo design~\cite{pluto}. Shared-PIM significantly outperformed prior work, demonstrating latency and energy improvements of $\sim$5$\times$ and $\sim$1.2$\times$ respectively when compared to the best existing solution, LISA. %\textcolor{blue}{
We further integrate Shared-PIM with pLUTo using DDR4 technology parameters for system-level evaluation. %} 
The integration of Shared-PIM with the pLUTo architecture resulted in performance gains of approximately 40\%, 44\%, 31\%, 29\% and 29\% for Matrix Multiplication (MM), Polynomial Multiplication (PMM), Number Theoretic Transform (NTT), Breadth-First Search (BFS) and Depth-First Search (DFS) benchmarks, respectively, while only increasing the area by 7.16\% compared to a pLUTo baseline. 
While we integrated Shared-PIM with pLUTo~\cite{pluto} in this work for evaluation purposes, Shared-PIM can be integrated with any DRAM-based PIM architecture such as \cite{sutradhar20233dl}, \cite{li2017drisa}, \cite{seshadri2017ambit}, etc.

% Our proposed Shared-PIM is a suitable candidate for increasing the efficiency of different PIM designs. Nevertheless, this paper exploits the integration of Shared-PIM to pLUTo\cite{pluto}, a state-of-the-art (SOTA) in-DRAM computing architecture. Our contributions are summarized into the following three points: \textbf{(1)} We introduce Shared-PIM, a feature that enables PIM arrays to simultaneously manage computation and data movement. \textbf{(2)} We introduce an inter-bitline shifting mechanism that can be used to move data between the columns of a array. \textbf{(3)} We developed a compiler for pLUTo that seamlessly integrates with NVBit, maximizing parallelism while accurately tracking data dependencies, ensuring optimized performance and efficient data flow within the architecture.

\section{Background and Related Work}
\label{sec:Background}
In this section, we briefly review the basic operation of a DRAM cell and the architecture of DRAM subarrays and banks. Furthermore, we review the MASA and pLUTo designs \cite{pluto, kim2012case}, which we integrate with Shared-PIM in Sec. \ref{sec:Evaluation}. Lastly, we provide an overview of existing strategies to mitigate the bottlenecks associated with data transfer in in-DRAM PIM architectures, specifically Rowclone \cite{seshadri2013rowclone} and LISA \cite{lisa}.

\subsection{DRAM}
\label{sec:Background_DRAM}

%\textcolor{blue}{
A modern DRAM organization typically involves multiple DRAM arrays organized in a hierarchical structure to optimize performance and density. A DRAM cell consists of a capacitor that stores the logical state in the cell, and an access transistor. The DRAM cell is connected to a sense amplifier, which detects and amplifies the logical state of the cell, resulting in a steady voltage level. Bitlines link the DRAM cells in one column to their respective sense amplifiers, whereas the wordlines control the cell's access transistor.%}

\begin{figure}[t]
    \centering
    \includegraphics[width=1.0\columnwidth]{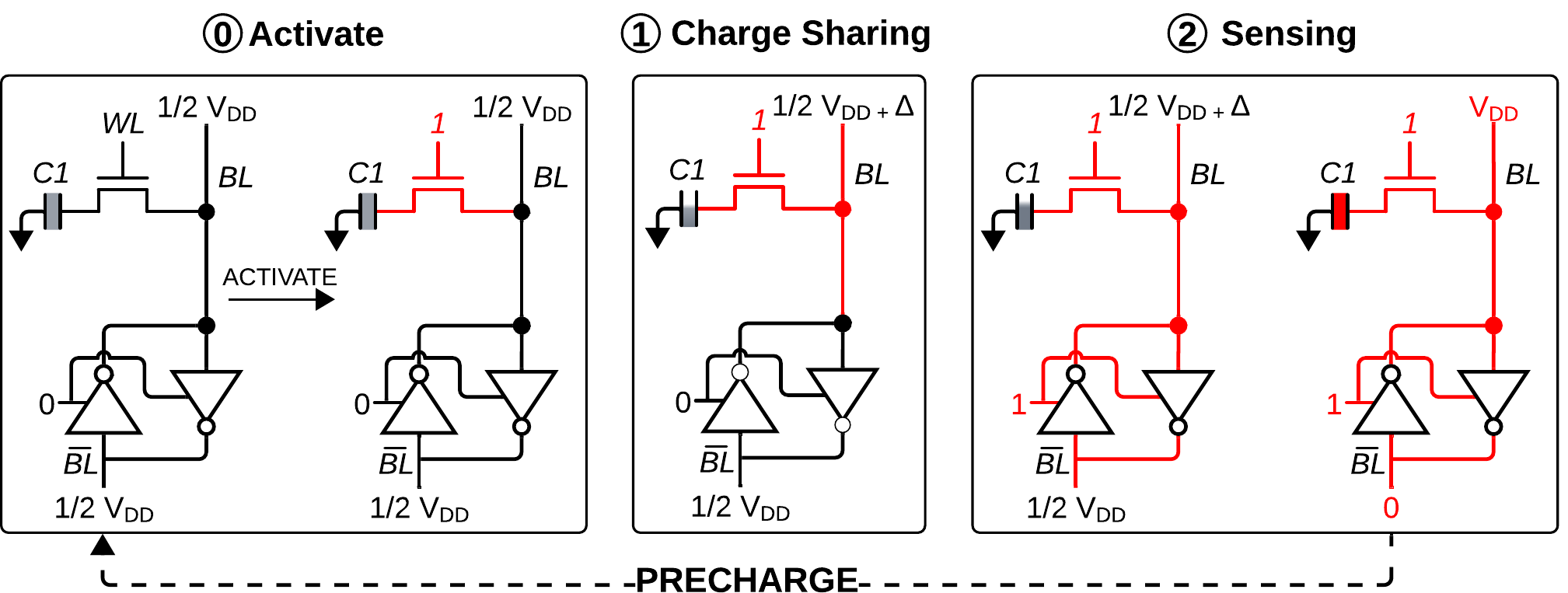} 
   
	\caption{The phases of read operation in a DRAM cell. (Red lines indicate an activated state)}
	\label{fig:DRAM_Op}
 
\end{figure}

%\textcolor{blue}{
A standard read operation in a DRAM cell involves three stages: \textcircled{0} \textbf{Activate}, \textcircled{1} \textbf{Charge Sharing}, and \textcircled{2} \textbf{Sensing} (Fig.\ref{fig:DRAM_Op}). Initially, the bitlines ($BLs$) are pre-charged to a $\frac{1}{2} V_{dd}$ state. After a read operation is initiated, an ACTIVATE signal is sent to the DRAM bank along with the source row address. The row decoders then enable the wordline ($WL$) corresponding to the address, thereby initiating the \textbf{Activate} phase \textcircled{0}. Once the $WL$ is enabled, the capacitor ($C_{1}$) in the DRAM cell is connected to the $BL$, and the \textbf{Charge Sharing} phase \textcircled{1} begins, which transfers charge between $C_{1}$ and $BL$. The charge on the $BL$ results in $\frac{1}{2} V_{dd}  + \Delta$ , where $\Delta$ can be positive or negative depending on whether the bit stored in $C_{1}$ is a `Logic 1' or a `Logic 0', respectively. When charge sharing is completed, the \textbf{Sensing} phase \textcircled{2} begins. Here, the sense amplifier is activated, which drives the $BL$ to a stable state based on the value of $\Delta$: When $\Delta$ is positive, the $BL$ is driven to $V_{dd}$, while the charge on $\overline{BL}$ goes to 0V. The charge on the $BL$ is reversed when $\Delta$ is negative \cite{keeth2007dram}. After the read operation is concluded and the $WL$ is deactivated, a PRECHARGE signal is sent to the bank to bring the bitlines back to the equilibrium state $(\frac{1}{2} V_{dd})$. %in preparation for the next operation.%}

Fig. \ref{fig:DRAM_Organization} depicts the standard DRAM architecture along with modifications required for our proposed Shared-PIM highlighted in red (these modifications will be discussed in Sec. \ref{sec:MMethodology}). In DRAM, memory is structured hierarchically:  (1) Ranks are logical or physical groupings of DRAM modules on a memory channel; (2) each rank contains multiple DRAM chips, each of which is further divided into banks; (3) banks are then divided into smaller subarrays per Fig. \ref{fig:DRAM_Organization}(a); (4) within each subarray, we reach the finest level of granularity -- i.e., the tiles, which contain memory cells per Fig. \ref{fig:DRAM_Organization}(b). While subarrays within a DRAM bank can function largely independently, they share a single global address latch. This shared resource necessitates sequential activation of subarrays within the same bank, limiting the potential for concurrent operations. 

% \textcolor{blue}{
The MASA mechanism, proposed in \cite{kim2012case}, exploits the inherent independence of DRAM subarrays to enable parallel activation of subarrays within DRAM architectures with minimal changes to the DRAM architecture. To support MASA, the memory controller monitors the status of all subarrays and rows within each bank to prevent issuing commands to already active subarrays, adding only minor storage overhead at the controller. Specifically, MASA incurs a DRAM chip area overhead of 0.15\% and a storage overhead of less than 256 bytes at the memory controller. In this work, we employ MASA to enable parallel activation of multiple subarrays within the same bank.
% }

Additionally, banks within a DRAM chip share a global row buffer, which supports I/O operations. 
Previous work \cite{seshadri2013rowclone} has demonstrated how the global row buffers can be used for inter-bank data transfer, as well as inter-subarray transfer by first copying the data to a temporary bank. However, as the global row buffer has a lower bit width compared to the bank row size, data must be copied serially/in smaller segments, thereby creating a bottleneck concerning inter-subarray data movement (as indicated by Rowclone's \cite{seshadri2013rowclone} PSM mode). Hence, no structure exists to facilitate data movement between subarrays in contemporary DRAMs, which negatively impacts PIM designs (i.e., AMBIT, pLUTo) that leverage standard DRAM organization.

% \subsection{pLUTO}

% \subsection{pLUTo}

% \begin{figure}[t]
%     \centering
%     \includegraphics[width=1\columnwidth]{./figures/Problem_statement_1.pdf} 
  
% 	\caption{\textcolor{blue}{Pipelining computation using Shared-PIM.}}
% 	\label{fig:pipelining}
 
% \end{figure}

pLUTo \cite{pluto} is a SOTA in-DRAM PIM architecture that leverages the high density afforded by DRAM to store large LUTs, together with newly added logic to perform in-DRAM LUT-based computation. In addition to the match logic, the pLUTo-BSA architecture extends the structure of standard DRAM by adding (a) a matchline and a transistor to every bitline in the subarray, along with (b) a buffer to store match results. When evaluating cryptographic and neural network compute workloads, pLUTo could offer speedups of 713$\times$ and energy savings of 1855$\times$ when compared to a CPU. pLUTo also outperforms other DRAM PIM designs by an average of 18.3$\times$, owing to its ability to handle parallel lookup queries directly in DRAM (i.e., pLUTo can efficiently manage complex operations while reducing data transfer demands).

% \textcolor{red}{pLUTo encounters several issues related to data transfer across computing arrays. The intra-operation dependencies in pLUTo arise from the limitations imposed by the size of LUTs. As these LUTs are utilized for basic computational units supporting only basic operations. For example, a simple 4-bit addition is straightforward, but to achieve a complex 32-bit addition, it necessitates the use of multiple 4-bit LUTs linked together. This arrangement requires pausing processing until data is transferred to the appropriate array, thereby incurring latency penalties. Moreover, inter-operation dependencies further exacerbate these issues, as data must be completely transferred between operations before subsequent computations can proceed, thereby increasing overall system latency.  }

%\textcolor{red}{
pLUTo requires extensive data transfers between subarrays during computations, which are managed by LISA \cite{lisa}. pLUTo encounters two distinct data transfer overhead scenarios during computations: one arises from the need to transfer data between LUTs across subarrays due to size limitations, and the other stems from the data transfer between operations within applications, such as in matrix multiplication.  The first type of overhead arises when performing complex operations, such as a 32-bit addition that cannot be contained within a subarray and instead requires the coordination of multiple 4-bit LUTs to support said addition that are distributed across subarrays. This leads to processing delays as data must be moved to the appropriate subarray. The second type of overhead occurs when computational steps are interdependent, necessitating a complete data transfer from previous operations before subsequent ones can proceed.  This requirement for sequential data handling, especially in application like matrix multiplication where multiplications and additions are depend on each other’s outputs. For example, in matrix multiplication, each product of matrix element pairs needs to be aggregated to compute the final matrix elements, resulting in data transfers accounting for approximately 60\% of the total operations involved in collecting and summing these results. The pLUTo architecture, utilizing the LISA framework, lacks the capability for concurrent computing and data transfer, resulting in significant system latency as operations must wait for previous data transfers to complete before proceeding.   %}

% \textcolor{red}{That said, the ability for pLUTo to process data dependencies represents a computational bottleneck which may in turn limit potential performance savings.} When data dependencies span multiple arrays, the computation must halt until data is transferred to the destination array, which can lead to latency penalties. Additionally, the LUT size is limited by the capacity of a single DRAM subarray. For more complex, higher-bit operations, the use of multiple DRAM subarrays in pLUTo becomes essential to implement large LUTs, which inevitably results in frequent inter-array data transfers.

% \subsubsection{Rowclone} 

% \textcolor{blue}{RowClone is a PIM architecture that exploits the high internal bandwidth available within DRAM to efficiently perform bulk data copy and initialization operations.RowClone introduces two DRAM operations: Fast Parallel Mode (FPM) and Pipelined Serial Mode (PSM). FPM facilitates the quick duplication of data from one row to another within the same DRAM subarray by issuing two consecutive activate commands. PSM allows for the sequential transfer of data across two banks via the shared internal bus. RowClone reduces the latency of a 4KB bulk copy operation by 11.6 times and cutting energy use by 74.4 times compared to conventional systems with only 0.01\% increase in DRAM chip area.  }

% \textcolor{blue}{Say something about the bad of Rowclone??}

\subsection{Related Work}
We briefly review existing strategies to mitigate the bottlenecks associated with data transfer in PIM architectures, specifically Rowclone \cite{seshadri2013rowclone} and LISA \cite{lisa}.

\subsubsection{Rowclone} 
\label{sec:rowclone}

Rowclone \cite{seshadri2013rowclone} allows for data movement within the same DRAM subarray by leveraging the inherent DRAM property that data in a DRAM cell is overwritten with the data (voltage level) on the bitline if the cell is connected to a bitline that is in a stable state (either $V_{dd}$ or $0$), as opposed to the equilibrium state $(\frac{1}{2} V_{dd})$. Rowclone can thus copy an entire row by (1) reading the source row onto the bitlines, (2) deactivating it, and (3) then sending an activate signal to the destination row (all while the sense amplifiers hold the data on the bitlines).
While Rowclone can efficiently copy data from a source row to any destination row within the same subarray, it does not directly address data movement between subarrays. To implement inter-array data transfer, Rowclone performs two bank-level Rowclone operations to move data to a temporary bank, and then back to the destination subarray.

\subsubsection{LISA} 
\label{sec:LISA_VS_SharedPim}

LISA~\cite{lisa} facilitates high-bandwidth connectivity between DRAM subarrays within the same bank, enhancing fast inter-subarray data movement. This is achieved by connecting neighboring subarrays with cost-effective isolation transistors that merge their bit lines. Using LISA, an activated row in a subarray can quickly transfer data between adjacent subarrays via a ``Row Buffer Movement" operation. LISA reports 26$\times$ higher bandwidth than traditional DRAMs. However, when integrated with existing PIM architectures (e.g., the pLUTo+LISA implementation described in \cite{pluto}), LISA does have limitations: (1) it cannot simultaneously support/perform data movement and computation, potentially reducing the throughput of PIM computations, (2) LISA's latency increases linearly as the distance between the source and destination subarrays increases, and (3) the number of stalled subarrays increases as the distance between arrays increases -- since LISA transfers data by linking all subarrays between the source and destination. In Sec. \ref{sec:MMethodology}, we explain in more detail how our Shared-PIM architecture can mitigate the aforementioned challenges.

% as depicted in Fig \ref{fig:LISA_vs_Shared-PIM}

\begin{figure}[t]
    \centering
    \includegraphics[width=1\columnwidth]{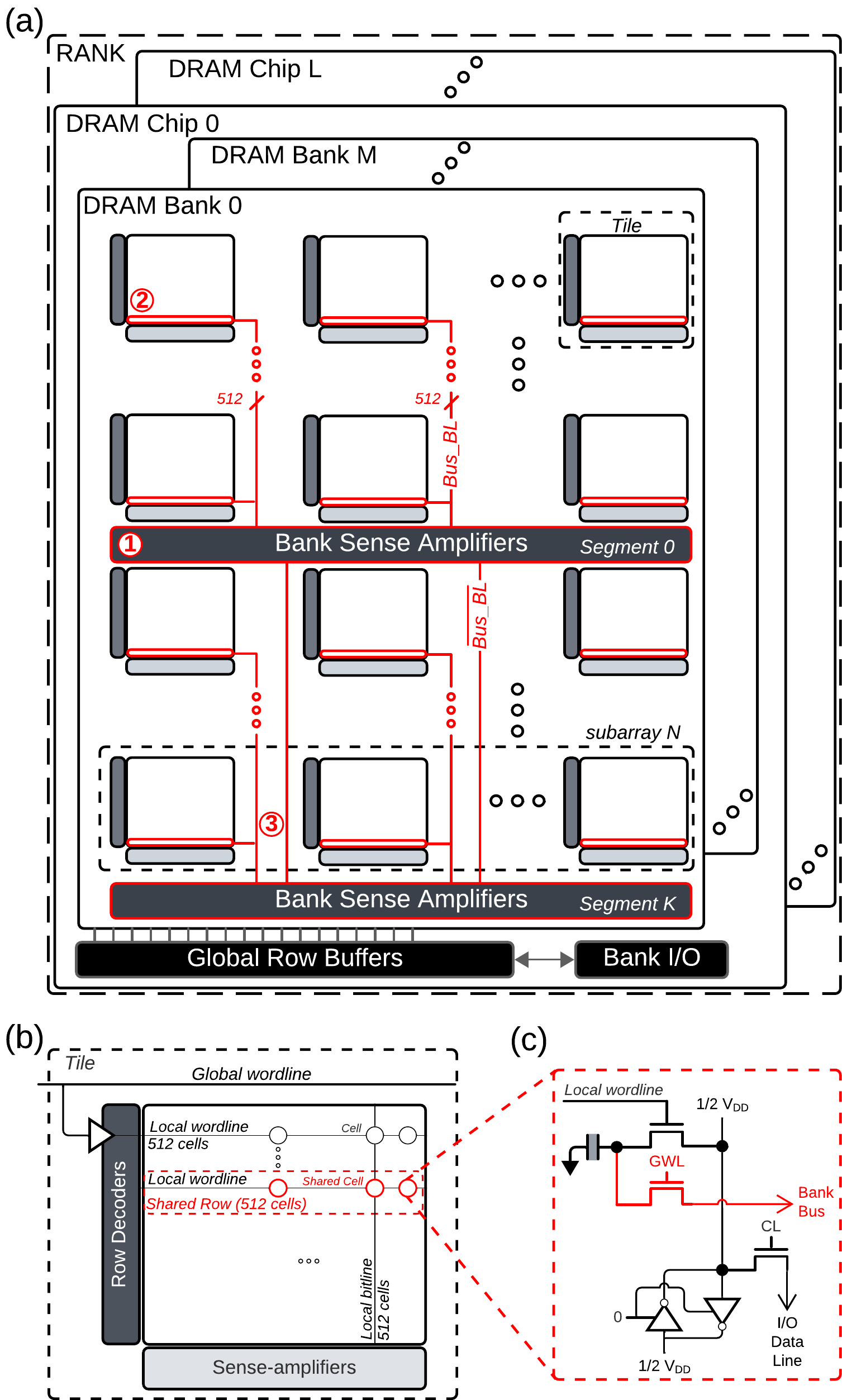} 
   % \vspace{-1ex}
	\caption{The Shared-PIM architecture: (a) The DRAM rank organization, (b) a tile of $512\times512$ DRAM cells, and (c) a single DRAM cell with an additional transistor, which forms a shared cell. (All parts highlighted in red in this figure are part of our proposed Shared-PIM architecture).}

 \label{fig:DRAM_Organization}
\end{figure}

\section{The Shared-PIM Architecture}
\label{sec:MMethodology}
This section presents Shared-PIM, our proposed architecture designed to streamline inter-subarray data transfers, which are a critical bottleneck for in-DRAM PIM systems.

\subsection{Architecture Overview}
\label{sec:Architecture}
 We propose Shared-PIM to reduce the latency of inter-subarray data movement in DRAM-based PIM solutions, aiming to efficiently manage unavoidable data dependencies. Our approach is rooted in the observation that when a subarray is active, its bitlines and sensing amplifiers are engaged. As a result, concurrent computation and data transfer between active subarrays is not possible without the presence of a separate set of bitlines/sense amplifiers to support such data movement. In the following, we discuss the physical structure of Shared-PIM and its implications to the in-DRAM PIM paradigm.

\subsubsection{Physical structure} 

Built on top of a standard DRAM organization, similar to other in-DRAM PIM designs like AMBIT and pLUTo, the proposed Shared-PIM design requires only minimal modifications to DRAM cells and peripheral circuitry (the overhead associated with these modifications will be discussed and evaluated in Sec. \ref{sec:Evaluation}). The changes incurred by Shared-PIM to the standard DRAM, highlighted in red in Fig. \ref{fig:DRAM_Organization}, specifically involve the introduction of the following key structures: 

\textbf{\textcircled{1}}~The \textit{bank sense amplifiers} (\textit{BK-SAs}), shown in Fig. \ref{fig:DRAM_Organization}(a), allow for sensing and reading data at the bank level while facilitating data movement between different subarrays within a bank. This setup supports concurrent data transfers and computation within a DRAM bank. 

\textbf{\textcircled{2}}~The \textit{Shared Row} (Fig. \ref{fig:DRAM_Organization}(b)), a row of augmented DRAM cells, which can be replicated multiple times in a subarray based on design choices. Augmented DRAM cells have dual access transistors for direct connection to the \textit{Bus\textunderscore BL} (the additional transistor is labeled as GWL in Fig. \ref{fig:DRAM_Organization}(c)). %\textcolor{blue}{
The combination of \textit{BK-SAs}, along with bank wide bitlines \textit{Bus\textunderscore BL}, creates a structure we refer to as a bank-level bus (\textit{BK-bus}). 

As indicated in Table \ref{tab:DRAM Configuration} in Sec.~\ref{sec:Evaluation}, we employ two \textit{Shared Rows} per subarray in this work, which allows one row to handle data transmission while the other is available for receiving data. This configuration is sufficient for most workloads.
That said, the optimal number of \textit{Shared Rows} in each subarray depends on the application. For instance, in scenarios where computations are simple and faster than data transfers, subarrays can compute partial results more quickly than the bus can transfer them. This causes the bus to become a bottleneck, forcing subarrays to stall until transfers are completed. Adding more shared rows could help mitigate this issue. Conversely, for workloads requiring more complex computations, additional shared rows would offer little benefit since most of them would remain idle for much of the time.

%\textcolor{blue}{As indicated in Table \ref{tab:DRAM Configuration} in Sec.~\ref{sec:Evaluation}, we employ two \textit{Shared Rows} per subarray in this work, which allows one row to handle data transmission while the other is available for receiving data. This configuration is sufficient for most workloads.
%} \textcolor{blue}{That said, the optimal number of \textit{Shared Rows} in each subarray depends on the application. For instance, in scenarios where computations are simple and faster than data transfers, subarrays can compute partial results more quickly than the bus can transfer them. This causes the bus to become a bottleneck, forcing subarrays to stall until transfers are completed. Adding more shared rows could help mitigate this issue but only to a limited extent, as the bus would eventually remain the primary bottleneck, leading to diminishing returns. Conversely, for workloads requiring more complex computations, additional shared rows would offer little to no benefit since most of them would remain idle for much of the time.
%}

% \textbf{(2)}~The \textit{bank-level bus} (\textit{BK-bus}), as wide as the subarray row, allows efficient inter-subarray data movement (Fig.

% From here on out we will refer to the combination of the BK-SAs and the Bus\textunderscore BL as one structure called the \textit{bank-level bus} (\textit{BK-bus}).

%\textcolor{blue}{
\textcircled{3} The \textit{BK-bus} is divided into shorter segments, each equipped with its own row of \textit{BK-SAs} and \textit{Bus\textunderscore BLs}. The segments of the \textit{BK-bus} are interconnected via their $\overline{Bus\textunderscore BLs}$, as illustrated in Fig. \ref{fig:DRAM_Organization}(a). Hence, when data is read onto one of the segments, the information is immediately transferred to all other segments through the $\overline{Bus\textunderscore BLs}$, acting as a unified structure. By breaking the \textit{BK-bus} into shorter segments, we mitigate potential latency issues due to the parasitic capacitance associated with long bitlines. 

To calculate the minimum number of segments required for the bus to operate at the desired latency, we first estimate the total parasitic capacitance of a long bitline that spans the entire bank vertically. We do this by using the DRAM internal block sizes and bitline capacitance information provided in Rambus's DRAM power and performance model \cite{rambus}. The number of bus segments is determined experimentally through SPICE simulations, conducted to identify the minimum number of segments that maintain the bus functionality. The selected number of BK-bus segments is listed in Table \ref{tab:DRAM Configuration}. Details concerning the SPICE simulations can be found in Sec. \ref{sec:Evaluation}.%}.

% \textcolor{blue}{Introducing \textit{Bus\textunderscore BLs} at the bank level could potentially lead to crosstalk noise, caused by parasitic coupling between adjacent bitlines in DRAM architectures, which has been well documented and studied in the literature \cite{keeth2007dram}. Studies have shown that bitline twisting can mitigate crosstalk \cite{al2004influence}, \cite{redeker2002investigation}. Other works, such as \cite{seyedzadeh2017mitigating}, propose a coding technique to mitigate row-based crosstalk. Since the \textit{Bus\textunderscore BLs} are regular bitlines, the same approaches can be applied to address potential issues with crosstalk noise in the proposed Shared-PIM architecture.}

% \textcolor{blue}{
Introducing \textit{Bus\textunderscore BLs} at the bank level may raise concerns regarding crosstalk noise, which arises from parasitic coupling between adjacent bitlines in general DRAM architectures (i.e., not exclusively in \textit{Shared-PIM}). The phenomenon has been well-documented and thoroughly studied in the literature \cite{keeth2007dram}. Research has shown that bitline twisting can effectively reduce or eliminate crosstalk between bitlines \cite{al2004influence}, \cite{redeker2002investigation}. Furthermore, other studies, such as \cite{seyedzadeh2017mitigating}, have proposed using a coding technique to mitigate row-based crosstalk. Since the \textit{Bus\textunderscore BLs} function as standard bitlines, these same strategies can be applied to address potential crosstalk noise issues in the proposed Shared-PIM architecture.
% }

% The Shared-PIM design utilizes the GWL transistors to connect cells in the shared row directly to the BK-bus; which is accessible to all subarrays within the bank. This configuration enables Shared-PIM to directly read data from the shared row onto the BK-bus. Similarly, writing data to the shared row from the BK-bus is possible, effectively bypassing the subarray's local sense amplifier. Consequently, Shared-PIM facilitates the movement of data from one subarray to another within the same bank, employing a method akin to~Rowclone (RC)-IntraSA~\cite{seshadri2013rowclone}. The process begins with a standard Rowclone operation to transfer data from the source row to the shared row using the two respective local wordlines within the same subarray. This is followed by copying the data across the BK-bus through two back-to-back activations of the GWL transistors in the shared row of the source and destination subarray. If the data is already stored in the shared row prior to computation, the process can be streamlined to a single copy operation over the BK-bus. 

\subsubsection{Shared-PIM vs. LISA} 

\begin{figure}[b]
    % \centering
    \includegraphics[width=1.0\columnwidth]{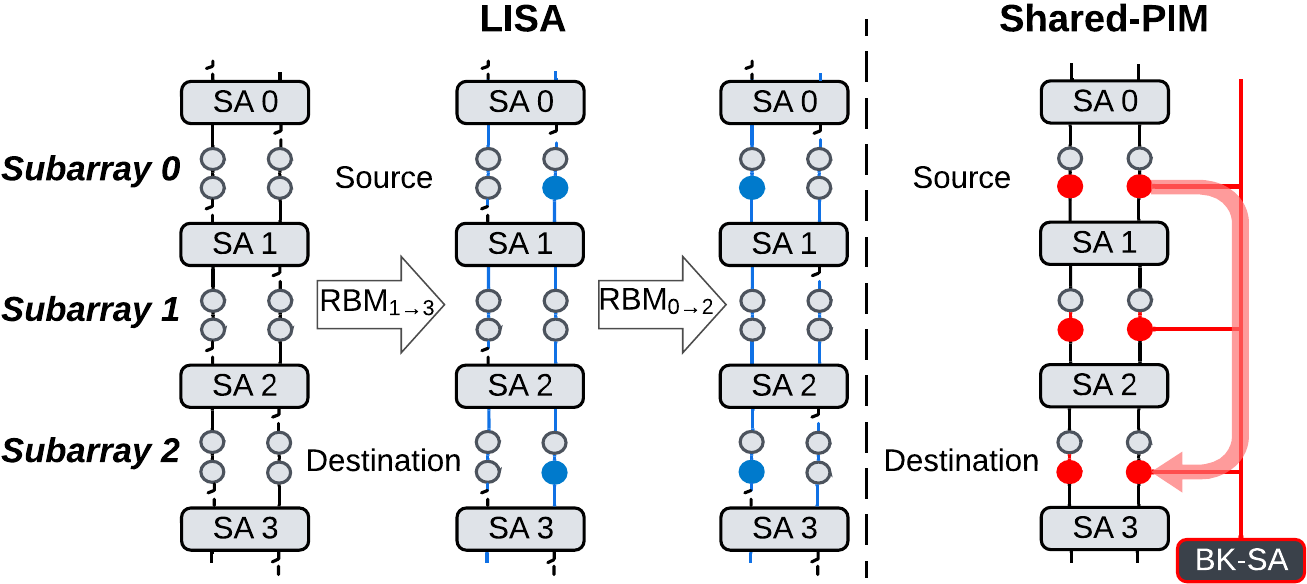} 
  
	\caption{Comparison of the inter-subarray copy mechanism of LISA \cite{lisa} versus Shared-PIM. Cells are represented by circles, and the bitlines/\textit{BK-bus} of Shared-PIM are represented by lines. The open-bitline structure \cite{takahashi2001multigigabit} is employed for both LISA and Shared-PIM. In this structure, two neighbouring subarrays share a common sense amplifier (SA).}
	\label{fig:LISA_vs_Shared-PIM}
 %\vspace{-2ex}
\end{figure}

\label{sec:Shared-PIM vs. LISA}
%\textcolor{blue}{
Fig.~\ref{fig:LISA_vs_Shared-PIM} depicts the differences in the copy mechanism between LISA and Shared-PIM. The DRAM cells are represented by circles. Blue circles/lines denote the activated cells/bitlines associated with each LISA operation. Meanwhile,  red cells/lines represent Shared-PIM's shared rows/\textit{BK-bus}.%}

%\textcolor{blue}{
LISA leverages the Row Buffer Movement (RBM) operation to perform inter-subarray data movement. The RBM operation links the bitlines of neighboring subarrays using isolation transistors. However, due to the open-bitline structure of the DRAM \cite{takahashi2001multigigabit}, LISA is unable to perform a complete row copy in one operation. Hence, LISA's copy mechanism is performed in two steps, copying one half of the row at a time. As illustrated in the figure, in order to copy one row of data from subarray 0 to subarray 2, LISA first activates the source row in \textit{subarray 0}, and then performs an RBM operation linking the bitlines of \textit{sense amplifier (SA) 1} to \textit{SA 3} ($RBM _{1\rightarrow3}$) that copies the first half of the source row to the destination. LISA then needs to perform a second RBM operation, which links the bitlines of \textit{SA 0} to \textit{SA 2} ($RBM _{0\rightarrow2}$), which completes the row copy operation.%}

%\textcolor{blue}{
Notably, LISA's copy mechanism must pass through subarray 1 in order to copy data from subarray 0 to subarray 2. This implies that all subarrays from the source and destination subarrays are stalled for the full duration of the copy operation. Additionally, with a single RBM operation, LISA can only move data up to one subarray away from the source by activating the links (isolation transistors) in their respective bitlines. This means that as the distance between the source and destination subarrays increases, the number of required RBM operations also increases, resulting in an increase in latency.%} 

%\textcolor{blue}{
In contrast, Shared-PIM utilizes the GWL transistors to connect cells in the shared row directly to the BK-bus, which is accessible to all subarrays within the bank (see Fig.~\ref{fig:DRAM_Organization}(c)). This configuration enables Shared-PIM to directly read data from the shared row onto the \textit{BK-bus}. Similarly, writing data to the shared row from the \textit{BK-bus} is possible, effectively bypassing the subarray's local SA. Consequently, Shared-PIM facilitates the movement of data from one subarray to another within the same bank, employing a method akin to~Rowclone (RC)-IntraSA~\cite{seshadri2013rowclone}. The process begins with a standard Rowclone operation that transfers data from the source row to the shared row, using the two respective local wordlines within the same subarray. This is followed by copying the data across the \textit{BK-bus} through two back-to-back activations of the GWL transistors in the shared row of the source and destination subarrays. If the data is already stored in the shared row prior to computation, the process can be streamlined to a single copy operation over the \textit{BK-bus}. We further discuss Shared-PIM's latency improvements and compare it against both RowClone \cite{seshadri2013rowclone} and LISA \cite{lisa} in Sec. \ref{sec: Circuit-level Latency, Energy, and Area Evaluation}.

\begin{figure*}[t]
    % \centering
    \includegraphics[width=2\columnwidth]{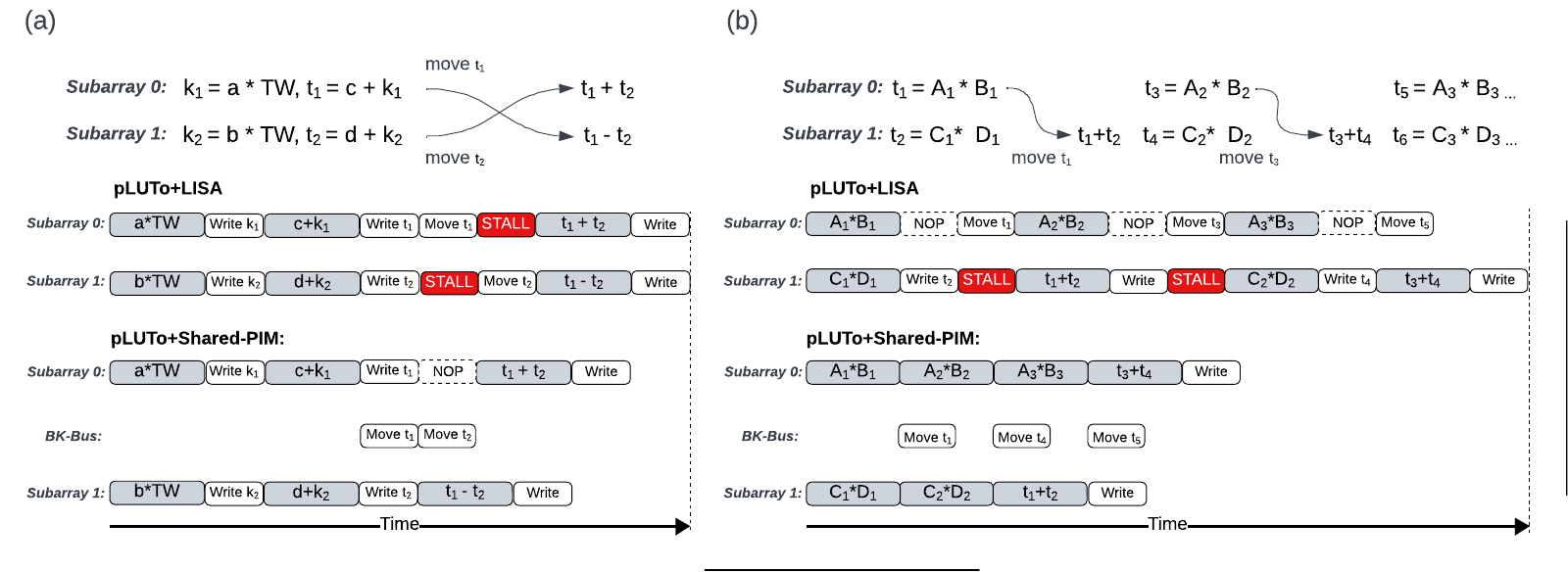} 
  
	\caption{%\textcolor{red}{
 (a) Pipeline example using Shared-PIM for a NTT butterfly computation. (b) Pipeline example using Shared-PIM for a matrix multiplication (commands not drawn to scale).}%}
	\label{fig:Application_Mapping}
 
\end{figure*}

\subsection{Memory Controller Support}
To support Shared-PIM and avoid conflicts, the memory controller must 1) recognize shared rows and 2) verify that neither address of a shared row is active before issuing an ACTIVATE command. As outlined in Section \ref{sec:Architecture} and shown in Fig.\ref{fig:DRAM_Organization}, shared rows have both local and global addresses (denoted by GWL), requiring that if one address is active, the other remains inactive until the operation completes. MASA \cite{kim2012case}, discussed in Sec. \ref{sec:Background}, is used in Shared-PIM to provide the memory controller with visibility into the active status of subarrays and rows, allowing for effective monitoring of shared row activity and prevention of conflicts. For the configuration in Table \ref{tab:DRAM Configuration} (1-channel, 1-rank, 4-chips, 4-banks per chip, 16-subarrays per bank), each subarray requires 11 bits to track its activation status, raised wordline, and column command designation. With a total of 256 subarrays in the system $(1 \times 1 \times 4 \times 4 \times 16)$, the overall storage overhead amounts to $256 \times 11 = 2816\ bits\ (352\ bytes)$. Hence, we estimate that the storage overhead required at the memory controller does not exceed 512 bytes.

\subsection{Application-Level Mapping}
\label{sec:Application_level_mapping}

\subsubsection{In-DRAM PIM considerations}

Our Shared-PIM architecture can move data from one subarray to another within the same bank without disturbing the source or the destination subarrays, thus allowing for concurrent computation and data transfer within the DRAM bank. This is because the copy operation is done over the BK-bus using the newly added \textit{BK-SAs}, which leaves the local SAs free to perform operations in parallel. This is critical to our overall objective of using the subarray as a processing element (PE). Furthermore, if subarrays are considered to be PEs for an in-DRAM PIM architectures, our proposed shared rows act as staging registers between the subarrays to achieve a pipelining effect, thereby improving the system's throughput.

Additionally, shared rows in different subarrays form a shared subarray when connected over the BK-bus, allowing for computation to be performed on data from different subarrays -- i.e., by performing triple activations on the bus as proposed in AMBIT \cite{seshadri2017ambit}. Finally, it is also possible to simultaneously copy data to multiple subarrays in a ``broadcasting” manner, i.e., by activating multiple destination rows simultaneously.  This has utility when the computational result(s) of one subarray is(are) needed in multiple other subarrays. SPICE simulations of the ``broadcasting” operation are presented in Sec. \ref{sec:Evaluation}.

% The benefits of the Shared-PIM architecture at the application level stem from the fact that computing subarrays do not need to wait for data transfers between subarrays to complete before they can perform with computations. This allows for more efficient processing as computation and data handling can occur simultaneously.

\subsubsection{Application-level use cases}
%\textcolor{red}{ 
To demonstrate the advantage of using Shared-PIM in conjunction with an existing in-DRAM PIM design (pLUTo), we present two application-level use cases in Fig \ref{fig:Application_Mapping}. We contrast our approach with the LISA architecture to highlight the specific improvements enabled by Shared-PIM.%}

%\textcolor{red}{
Fig.~\ref{fig:Application_Mapping}(a) illustrates the mapping of the butterfly computation step in the Number Theoretic Transform (NTT) using two subarrays. In the figure, `a' and `b' represent the initial input data points, where 'a' is processed in subarray 0 and `b' in Subarray 1, with each being multiplied by the twiddle factor `TW' to produce intermediate results for further computations. In the \textbf{pLUTo + LISA configuration}, a significant delay is observed because each subarray must wait until data transfer is complete before continuing with further computations, as marked by $STALL$. While the Shared-PIM architecture also involves a wait for the \textit{BK-bus} to complete $Move_t{1}$ before progressing with further computations due to data dependencies, this waiting time is not considered a stall because the subarray is available to be used for other computations during this period. This interval is labeled $NOP$, standing for non-operation in the subarray. Despite the $NOP$, the latency of the Shared-PIM for this butterfly computation is lower than that of LISA.%}

%\textcolor{red}{
Fig \ref{fig:Application_Mapping}(b) shows the mapping of a matrix multiplication segment using two subarrays. Here, `A1', `A2', `A3', `B1', `B2', and `B3' are the matrix elements handled by subarray 0, and `C1', `C2', `C3', `D1', `D2' and `D3' by subarray 1, with each pair being multiplied to contribute to the final matrix product. Subarray 0 is tasked with the operations \( A_i \times B_i \), and Subarray 1 with \( C_i \times D_i \). Once \( A_1 \times B_1 \) and \( C_1 \times D_1 \) are computed, the results $t_{1}$ and $t_{2}$ are immediately moved to the \textit{BK-bus}, which is indicated by $Move t_{2}$. This setup allows the subarrays to continue with subsequent operations like \( A_2 \times B_2 \) and \( C_2 \times D_2 \) without waiting for the data transfer, thereby overlapping data movement with computation. Furthermore, as \( A_2 \times B_2 \) does not depend on \( t_{1} + t_{2} \), in the pLUTo+Shared-PIM configuration, subarray 0 can immediate compute \( A_2 \times B_2 \) after \( A_1 \times B_1 \). In contrast, in the pLUTo+LISA setup, both subarrays 0 and 1 are occupied with transferring $t_{1}$ so they cannot immediately perform any subsequent computation.%}

%\textcolor{red}{
It is important to highlight that the data transfer time, denoted as $Move$ in Fig \ref{fig:Application_Mapping}, is shorter in Shared-PIM compared to LISA since LISA's copy operation involves two separate steps as explained in Sec.\ref{sec:Shared-PIM vs. LISA}. Moreover, Shared-PIM also consumes less energy during data transfers. The primary reason for these advantages is that Shared-PIM utilizes a direct link between subarrays, enabling data to be copied from the source row to the destination row in a single operation (see Sec \ref{sec:LISA_VS_SharedPim}). We compare our Shared-PIM approach with LISA in Sec. \ref{sec:Application_level_mapping}.%}

\section{Evaluation}
\label{sec:Evaluation}
In this section, we verify the functionality and assess the performance of Shared-PIM (integrated with pLUTo \cite{pluto}) at both the circuit and application levels using DDR3 and DDR4 technology parameters, respectively. At the circuit-level, we compare Shared-PIM's copy operation directly against RC-InterSA~\cite{seshadri2013rowclone} as well as LISA~\cite{lisa}. At the application-level, we compare the pLUTo+Shared-PIM design with the pLUTO+LISA design~\cite{pluto}.

\subsection{Experimental Setup}
% \label{sec:exp_setup}
Here, we explain the setup used to assess Shared-PIM's functionality at the circuit-level, as well as to evaluate its performance at the application-level. 

\subsubsection{Circuit level}
\label{sec:exp_setup}
We implemented the Shared-PIM subarrays in SPICE using ASU's low power 45nm predictive technology model (PTM) \cite{zhao2007predictive}, with the JEDEC DDR3-1600 (11-11-11) timing parameters \cite{jesd792012ddr3}. Note that this is the same setup used by LISA. Furthermore, we evaluated the latency and energy of Shared-PIM using the DDR3 DRAM configuration shown in Table \ref{tab:DRAM Configuration}. We opted to include two shared rows per subarray so that one is always available for facilitating data transfers, while the other is utilized to store the computation result that will be transferred on the following cycle. 

Our simulations confirmed that Shared-PIM can operate with the JEDEC DDR3-1600 (11-11-11) timing parameters, even with our modifications at the bank level (i.e., the addition of the BK-bus and BK-SAs), and the extra transistor in augmented DRAM cells. For energy evaluations, we employ the power models from Micron \cite{micron} and Rambus \cite{rambus} to estimate the power consumption of the DRAM commands; these results were multiplied by the latency to quantify the energy dissipation of a copy operation using Shared-PIM. We estimate the Shared-PIM area based on the DRAM area breakdown reported in \cite{pluto}, along with added interconnect and transistor count overhead (the same method used in \cite{pluto}).

\begin{table}[t]
\caption{The DRAM configuration used in Shared-PIM}
\label{tab:DRAM Configuration}
\resizebox{\columnwidth}{!}{
\begin{tabular}{cl}
\hline
\textbf{Model} &
  \multicolumn{1}{c}{\textbf{Configuration}} \\ \hline
\begin{tabular}[c]{@{}c@{}}DDR3-1600 \\ (11-11-11)\end{tabular} &
  \begin{tabular}[c]{@{}l@{}}533 MHz, 8 GB, 1-channel, 1-rank, 4-chips, 4-banks per chip,\\ 16-subarrays per bank, 4 BK-bus segments,\\ 512 rows per subarray, 2-Shared rows per subarray, 8KB per row.\end{tabular} \\ 
  \\
\begin{tabular}[c]{@{}c@{}}DDR4-2400T \\ (17-17-17)\end{tabular} &
  \begin{tabular}[c]{@{}l@{}}2400 MHz, 8 GB, 1-channel, 1-rank, 4-chips, 4-banks per chip,\\ 16-subarrays per bank, 4 BK-bus segments,\\ 512 rows per subarray, 2-Shared rows per subarray, 8KB per row.\end{tabular} \\   
  \hline
\end{tabular}
}
\end{table}

\subsubsection{Application level}

% To evaluate the application-level performance of a pLUTo~+~Shared-PIM design, we compiled simulation data from Shared-PIM—including latency and energy metrics -- and combined said results with the computational performance reported from pLUTo paper. This aggregated dataset, along with the instruction list compiled from the modified NVBit, was then input into a cycle-accurate simulator, enabling us to analyze the overall performance of the integrated pLUTo~+~Shared-PIM design.
To evaluate the application-level performance of the pLUTo+LISA and the pLUTo+Shared-PIM designs, we employed the inter-subarray data transfer latency and energy results reported in \cite{lisa}, as well as our results from Shared-PIM, and combined them with the computational performance of pLUTo reported in \cite{pluto}. Importantly, since \cite{pluto} used the ASU's low power 22nm PTM \cite{zhao2007predictive} and the DDR4 timing standards, %\textcolor{blue}{
we first employ SPICE simulations to test Shared-PIM using the DDR4-2400T (17-17-17) timing parameters used by \cite{pluto}. %} We then scaled the latency and energy results of LISA and Shared-PIM using the JEDEC DDR4-2400T (17-17-17) timing parameters \cite{jesd792017ddr4} and the method presented in \cite{stillmaker2017scaling}, so they are consistent with the setup utilized by pLUTo.

Finally, the combined computational and data transfer performance of pLUTo+LISA and pLUTo+Shared-PIM were input into a Python-based, cycle-accurate simulator that provides a detailed cycle-by-cycle analysis of computation and subarray utilization. Notably, there is generally good agreement between our results and the results reported in \cite{pluto}. Our method of evaluation, which employs (1) the cycle-accurate simulator, (2) the data transfer time of LISA, and (3) the operation latency (e.g., addition, multiplication, etc.) for end-to-end tasks (e.g., CNN inference) reported in pLUTo, is consistently within 15\% of the performance reported in \cite{pluto}.

%However, it's essential to emphasize the primary objective of this paper: mitigating the inter-array data transfer overhead in PIM designs. Our primary focus isn't to achieve an exact replication of pLUTo's baseline performance but to demonstrate the potential improvements offered by Shared-PIM. By comparing the performance of pLUTo with Shared-PIM to that of pLUTo with LISA (our baseline), using our evaluation methodology, we aim to underscore the tangible benefits of our design. The essence lies not in the absolute correctness of pLUTo's baseline performance but in the relative enhancement that Shared-PIM brings to the table.

\subsection{Circuit-level Functionality Validation}

\begin{figure}[b]
    \centering
    \includegraphics[width=1.0\columnwidth]{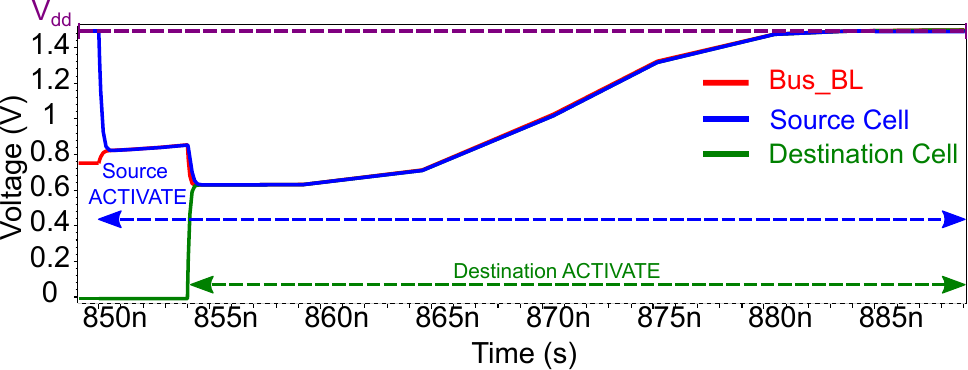} 
  
	\caption{A SPICE simulation of Shared-PIM transferring data from a source row to 4 destination rows in different subarrays.}
	\label{fig:SPICE}
 
\end{figure}

Fig.~\ref{fig:SPICE} depicts the waveforms from our SPICE simulations of Shared-PIM to simultaneously copy one source row to four destination rows in different subarrays over the \textit{BK-bus} (i.e., a ``broadcasting" operation). Once the source row is activated, the charge sharing stage between the cells and the \textit{BK-bus} begins. At the beginning of the operation we observe the voltage of the source cell (\textbf{blue line}) declining, while the voltage on the \textit{BK-bus} bitline (\textbf{red line}) rises as they share charge. Once the cell and the bitline reach equilibrium, the \textit{BK-SAs} start sensing and amplifying the voltage on the bitline. We then activate the destination row(s) $4ns$ later, which causes the voltage drop at $854ns$, while the voltage in destination cell (\textbf{green line}) rises. Similarly, after the charge sharing stage is finished, the \textit{BK-SAs} begin amplifying the voltage on the bitline until the data is restored into the source and destination rows, which concludes the data transfer operation. %\textcolor{blue}{
Although four destination rows is not an absolute limit for the broadcast operation, and our simulations show that broadcasting to five or even six destination rows is possible, we keep the maximum number of destination rows to four to ensure it stays within the standard DDR timing limits.%}

\subsection{Circuit-level Latency, Energy, and Area Evaluation}
\label{sec: Circuit-level Latency, Energy, and Area Evaluation}

\begin{figure}[t]
    \centering
    \includegraphics[width=1\columnwidth]{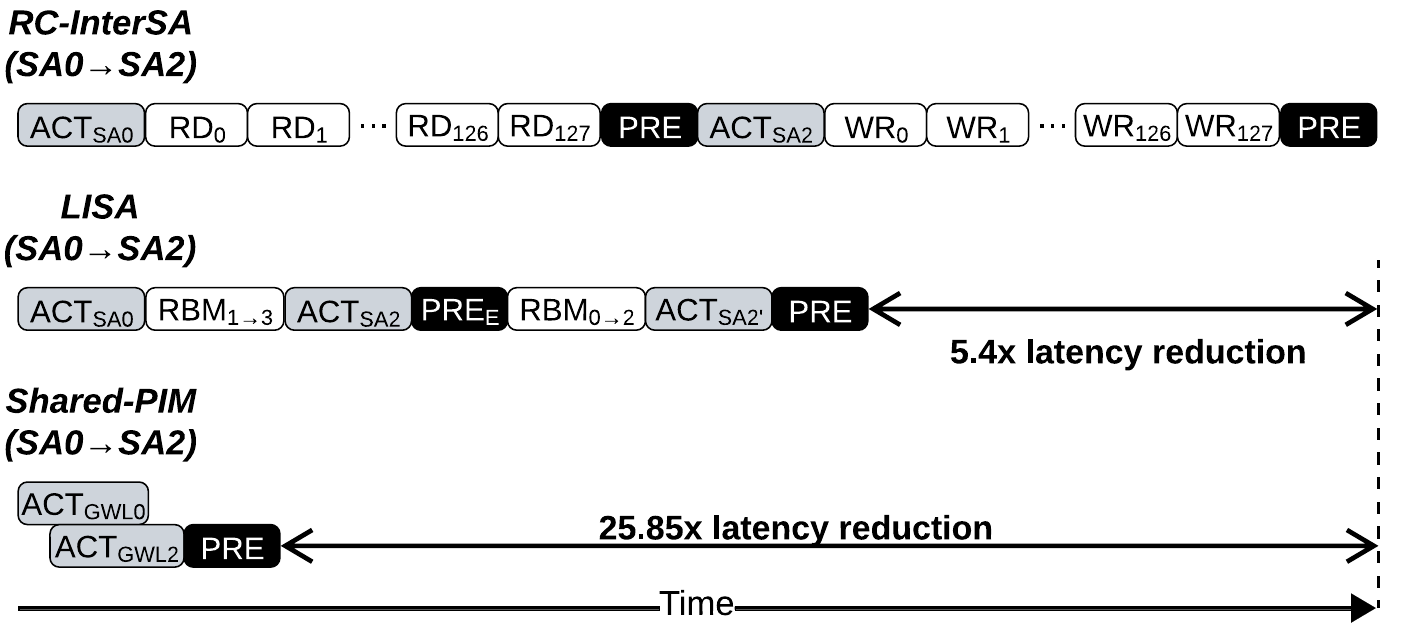} 
  
	\caption{Shared-PIM command timeline compared with LISA-RISC and RC-InterSA (commands not drawn to scale).}
	\label{fig:timeline}
 
\end{figure}

Table \ref{tab:Latency} reports the inter-subarray latency and energy for copying one row (8KB of data) from one subarray to another within the same bank using four different approaches: (1) memcpy, which uses the memory channel, (2) Rowclone (RC-InterSA), (3) LISA, and (4) our proposed Shared-PIM (with two shared rows per subarray). %As described in section \ref{sec:MMethodology}, copying an entire row from a shared row of a subarray to another subarray requires one Rowclone operation over the BK-bus. The Rowclone operation includes two ACTIVATE operations and one PRECHARGE operation, incurring a copy latency of 83.75ns, which follows the DDR3-1600 (11-11-11) timing parameters. 

% With respect to intra-bank copy latency, the BK-bus in Shared-PIM is connected to a total of 32 shared rows, thus the capacitance on the bus bitlines is lower than that of the a 512-row subarray. Hence, our BK-bus can operate at a faster latency compared to regular subarrays. LISA reported that a 32 row subarray can operate at the following reduced timing parameters: tRCD=7.5ns, tRP=8.5ns, and tRAS=13ns. Our SPICE simulations confirmed that our BK-bus can also operate within the same reduced timing parameters. 

With respect to inter-subarray copy latency, we further optimize Shared-PIM's copy latency by overlapping the two ACTIVATE commands with only a 4ns delay between them as proposed in \cite{seshadri2017ambit}, resulting in a total copy latency of $52.75ns$. Fig.~\ref{fig:timeline} shows the command timeline and latency improvement of Shared-PIM with respect to RC-InterSA \cite{seshadri2013rowclone} and LISA \cite{lisa}. 

%We employ the power models by Micron \cite{micron} and Rambus \cite{rambus} to estimate the power consumption of the DRAM commands, and multiply that by the latency to get the energy dissipation of a copy operation using Shared-PIM. 
The energy consumption reported in Table \ref{tab:Latency} was obtained with the method described in \ref{sec:exp_setup}, i.e., by calculating the power usage of the DRAM commands, and then multiplying this by the latency of a copy operation via Shared-PIM. Notably, the addition of two shared rows per subarray does not significantly affect the subarray's energy, since it only adds two transistors per bitline (16K transistors per subarray, in total). Meanwhile, the addition of the \textit{BK-SAs} that enable Shared-PIM's data transfer raises the power consumption per copy operation. %\textcolor{blue}{
However, because the copy process is performed on the BK-bus with improved latency, the overall energy dissipation per copy operation is minimized compared to Rowclone \cite{seshadri2013rowclone} and the conventional memcpy. When comparing against LISA~\cite{lisa}, it is apparent that Shared-PIM's energy savings (1.2$\times$) are not as good as the latency improvements (5$\times$). This can be attributed to Shared-PIM's additional \textit{BK-SAs}, since the copy operation over the bus activates the entire bus which consists of four rows of sense amplifiers as explained in Sec.~\ref{sec:MMethodology}, i.e., Shared-PIM's copy operation activates four times the number of sense amplifiers that LISA does, which results in a trade off of power dissipation for latency improvements.%}
 %Table \ref{tab:Latency} shows Shared-PIM latency and energy measurements compared to other data transfer solutions.

Finally, Table \ref{tab:Area} shows the area breakdown of the base DRAM, the pLUTO architecture (using the BSA mode, which is pLUTo's more compact implementation), and Shared-PIM (combined with pLUTo). We estimate a 7.16\% overhead to pLUTO's chip area with Shared-PIM, which is a result of the addition of BK-bus, BK-SAs, the GWL transistors (for shared cells), and GWL drivers. To further reduce the physical implementation complexity and the area overhead of Shared-PIM, the \textit{BK-bus} lines could be implemented using a different metal layer than that of the regular bitlines.

\begin{table}[t]
\caption{Inter subarray copy latency and energy}
\label{tab:Latency}
\resizebox{\columnwidth}{!}{
\begin{tabular}{lll}
\hline
\multicolumn{1}{c}{\textbf{Copy Commands (8KB)}} & \multicolumn{1}{c}{\textbf{Latency (ns)}} & \textbf{Energy ($\mu$J)} \\ \hline
\multicolumn{1}{c}{memcpy (via mem. channel)} & 1366.25               & 6.2                \\
RC-InterSA                                      & 1363.75               & 4.33               \\
LISA                    & 260.5 & 0.17 \\
\textbf{Shared-PIM}                           & \textbf{52.75}         & \textbf{0.14}      \\ \hline
\end{tabular}
}
\end{table}

\bgroup
\begin{table}[t]
\caption{Area Overhead Comparison}
\label{tab:Area}
\resizebox{\columnwidth}{!}{
\begin{threeparttable}

% \resizebox{\columnwidth}{!}{
\begin{tabular}{cccc}
\hline
\textbf{}          & \multicolumn{3}{l}{\textbf{Area ($mm^2$)}}                                                                                  \\ \hline
\textbf{Component} & \textbf{BASE DRAM} & \textbf{pLUTo-BSA}                                        & \textbf{\begin{tabular}[l]{@{}c@{}}pLUTo+\\ Shared-PIM\end{tabular}}                                       \\ \hline
DRAM cell                & 45.23 & 45.23 & 45.29 \\
Local WL driver          & 12.45 & 12.45 & 12.45 \\
Match logic              & -     & 4.61  & 4.61  \\
Match lines              & -     & 0.02  & 0.02  \\
Sense amp                & 11.40 & 18.23 & 18.23 \\
Row decoder              & 0.16  & 0.47  & 0.47  \\
Column decoder           & 0.01  & 0.01  & 0.01  \\
GWL driver               & -     & -     & 0.05  \\
BK-bus lines                & -     & -     & 0.04  \\
BK-SAs   & -     & -     & 5.70  \\
Shared-PIM Row decoder & -     & -     & 0.01  \\
Other                    & 0.99  & 0.99  & 0.99  \\
\textbf{Total}     & \textbf{70.24}              & \begin{tabular}[c]{@{}c@{}}\textbf{82.00}\\ \end{tabular} & \begin{tabular}[c]{@{}c@{}}\textbf{87.87}\\ \textbf{(+7.16\%)}\tnote{1}\end{tabular} \\ \cline{1-4}

\end{tabular}

% }
\begin{tablenotes}
       \item [1] Shared-PIM's overhead \% is calculated with respect to pLUTo.
\end{tablenotes}
\end{threeparttable}
}
\end{table}
\egroup

\subsection{Application-level Evaluation}
\label{sec:system_level_eval}

\begin{figure}[t]

  \centering
  \resizebox{\columnwidth}{!}{\includegraphics{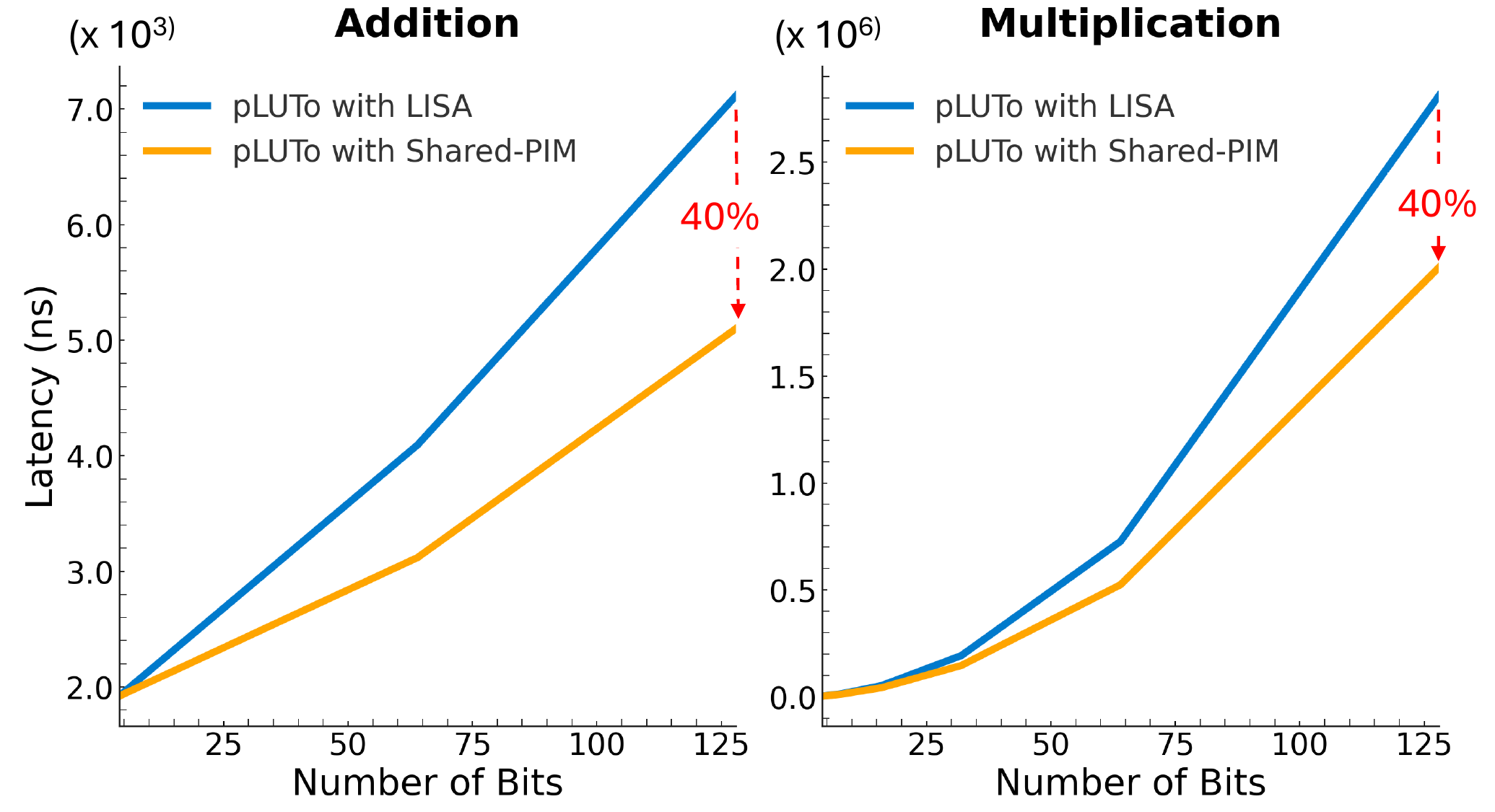}}

  \caption{Latency comparison between pLUTo with LISA and pLUTo with Shared-PIM for addition and multiplication operations with varied bit sizes.
  }
  \label{fig:Higher_Bits_Comparison}
% \vspace*{-3mm}
\end{figure}

\begin{figure}[t]

  \centering
 \includegraphics[scale=0.27]{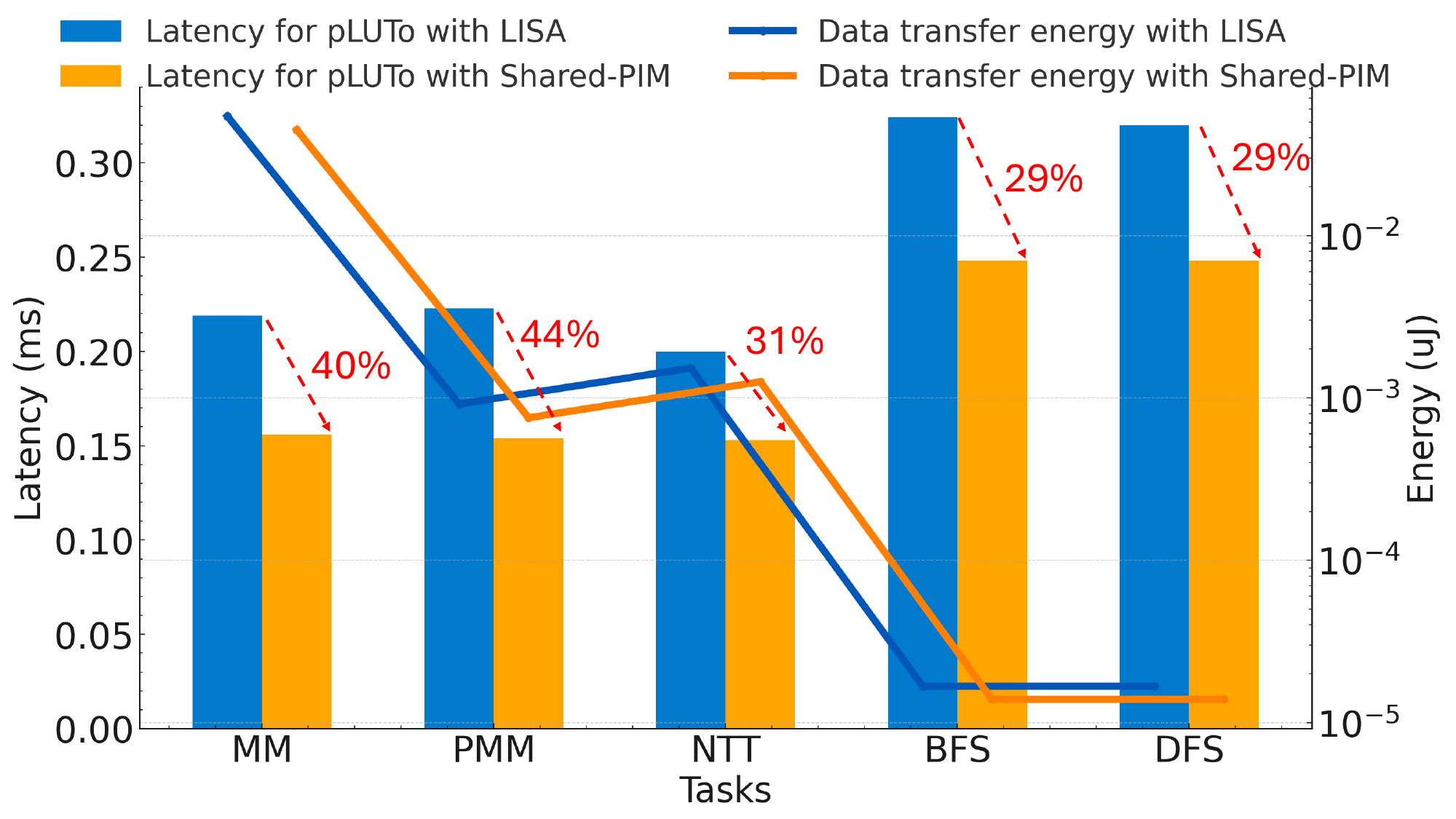}

  \caption{Comparison of execution time and data transfer energy consumption between pLUTo with LISA and pLUTo with Shared-PIM for five applications.
  }
  \label{fig:Applications}
% \vspace*{-5mm}
\end{figure}

Here, we evaluate how data transfer in two architectures influences  application-level performance. The architectures considered are pLUTo combined with LISA \cite{pluto}, and a scenario where pLUTo is integrated with our proposed Shared-PIM. We carry out our evaluation on two different fronts: (1) with operations of varied number of bits, and (2) by running five application benchmarks.

For varied number of bits, we study pLUTo's performance for addition and multiplication operations. Given the size constraints of the LUTs in pLUTo, a single subarray cannot accommodate the entire LUT required for computations with higher number of bits. As a result, these computations are distributed across multiple subarrays. This distribution necessitates inter-subarray data transfers when executing a single operation, which can impact performance. In our experiments, we consider that a single pLUTo subarray can efficiently perform a 4-bit addition and 4-bit multiplication as reported in \cite{pluto}. We then test pLUTo's performance across varying bit-widths:  16, 32, 64 and 128 bits for both addition and multiplication. The inclusion of 128-bit operations demonstrates the advantages of using Shared-PIM in pLUTo for applications requiring high-precision computing and cryptography.  Throughout these experiments, we assume that pLUTo executes these computations with maximum parallelism. This assumption of maximum parallelism aims to mimic an environment with optimal bank-level parallelism, allowing us to demonstrate the benefits of Shared-PIM under high level optimization.

Fig.~\ref{fig:Higher_Bits_Comparison} compares the latency of pLUTo paired with LISA and pLUTo combined with Shared-PIM, for addition and multiplication, with varied number of bits. The benefits of Shared-PIM become increasingly apparent in computations involving operations with higher number of bits, as they require more data movement.

The speedup of Shared-PIM for addition operations can be attributed to its capability to execute all the 4-bit additions simultaneously. After these parallel operations, the results are forwarded to a subarray for final aggregation via the \textit{BK-bus}. Shared-PIM enables efficient data transfer, allowing it to achieve a 40\% performance improvement in 128-bit addition compared to LISA.

On the other hand, the scenario for multiplication is more challenging. Data dependencies in multiplication, especially for operations with higher number of bits, make it more challenging to achieve the same level of parallelism as in addition. This complexity stems from the combination of 4-bit multiplication, addition, and shifting. However, Shared-PIM offers significant advantages by enabling simultaneous computation and data transfer. For instance, while intermediate multiplication results are being transferred for final aggregation, Shared-PIM allows the next layer of multiplication and shifting operations to proceed immediately. Shared-PIM paired with pLUTo leads to a 40\% improvement in performance for 128-bit multiplication compared to pLUTo combined with LISA.

%While the efficiency of operations such as addition and multiplication is crucial, it is also essential to see how pLUTo combined with Shared-PIM performs in real-world application scenarios. To this end, 

% For application-level study, we explored the performance impact of our Shared-PIM design on pLUTo across three applications, each with varying data dependencies. 

% The 7-layer CNN Inference processes 224x224 images with 64 channels using distinct 3x3 filters for each layer, which necessitates significant data transfer between layers, albeit with low data dependencies. The FFT application transforms a signal into its frequency components over 1000 data points, with pronounced data transfer in the Cooley–Tukey computation step. In contrast, the BFS application analyzes a 1000-node densely connected graph, where each node links to every other node, marking the highest amount of data dependencies among the trio. Notably, all computations in these applications utilize 32-bit operations, with latency and energy derived from previous tests for pLUTo with LISA and Shared-PIM.

%\textcolor{red}{
For the study with application benchmarks, we explored the performance impact of our Shared-PIM design on pLUTo across five programs, which were selected for their varying levels of data dependencies. This selection allows us to compare and contrast the performance benefits of using the Shared-PIM architecture across different dependency scenarios. The benchmark programs are: %}

\begin{enumerate}

\item Number Theoretic Transform (NTT): The NTT transforms polynomials into their modular equivalents in a finite field, functioning similarly to the  Fast Fourier Transform (FFT). We evaluate the NTT using polynomial degree of 300.

\item Breadth-First Search (BFS): The BFS application analyzes a 1000-node densely connected graph, where each node links to every other node, marking the highest amount of data dependencies among all of the selected programs.

\item Depth-First Search (DFS): The DFS is similar to BFS, being a worst-case DFS that needs to visit each node and explore its neighbors.

\item Polynomial Multiplication (PMM): Polynomial multiplication is the process of multiplying two polynomials to obtain their product. We use polynomials with a degree of 300. We employ a naive approach to polynomial multiplication (not using NTT to optimize).

\item Matrix Multiplication (MM): Two matrices with a size of 200 $\times$ 200 multiply each other. MM has a high amount of data dependencies and requires significant data transfer between the elements of the matrices.

\end{enumerate}

%\textcolor{red}{
All the computations in these benchmark programs use 32-bit operations, and the latency and energy consumption are derived from previous tests for pLUTo with LISA and Shared-PIM architectures. We assume that both pLUTo with LISA and pLUTo with Shared-PIM can run these applications under full parallelism, which in this context means that an ideal number of computing arrays are available for these benchmark programs. %}

%\textcolor{red}{
Fig.~\ref{fig:Applications} presents the latency and energy consumption metrics for the benchmarks executed under pLUTo with LISA and the Shared-PIM architectures. The latency comparison reveals that Shared-PIM confers a substantial speed advantage in all tasks, with a 40\% latency reduction in MM, 44\% in PMM, and 31\% in NTT. Shared-PIM enhances performance by enabling concurrent computation and data transfer within pLUTo subarrays, which facilitates the pipelining of data transfer with computation processes. Additionally, as shown in Fig.\ref{fig:Higher_Bits_Comparison}, pairing Shared-PIM with pLUTo achieves an 18\% speedup in 32-bit addition and a 31\% speedup in 32-bit multiplication compared to the pLUTo with LISA setup, significantly boosting performance for individual operations. 

The speedup for MM and PMM are particularly significant, as they are attributed to similar degrees of parallelism and computational steps. Compared to NTT, MM and PMM demonstrate greater speedup. This is primarily due to relative lower data dependencies in MM and PMM compared to NTT, allowing Shared-PIM to more effectively parallelize data transfer and computation in MM and PMM (as discussed in Sec.~\ref{sec:Application_level_mapping}). In our analysis, BFS and DFS show equal performance in worst-case scenarios on the same graph, as they follow identical processes.%}

%\textcolor{red}{
With respect to energy consumption, a clear difference emerges despite the parallel capabilities of both designs. The Shared-PIM architecture proves to be more energy-efficient in data transfers than LISA, as indicated by the consistent reduction in energy usage. As a result, Shared-PIM achieves an average of 18\% energy savings in data transfers across all benchmarks.% }

%\textcolor{blue}{Non-PIM situation}
\subsection{Evaluation of non-PIM Scenarios}

\begin{table}[t]
\centering
\caption{Non-PIM Simulation settings}

\label{tab:Sim_para}
\begin{tabular}{l|l}
\hline
\multicolumn{2}{c}{\textbf{Processor Parameters}}                    \\ \hline
Core            & Single Core, X86, OoO, 3GHz                \\ \hline
L1 Cache        & 10 Cycles, 32KB, 2-Way                      \\ \hline
L2 Cache        & 20 Cycles, 256KB, 8-Way                      \\ \hline
LLC             & 30s Cycles, 8MB, 16-Way                    \\ \hline

\multicolumn{2}{c}{\textbf{Memory Parameters}}          \\ \hline
Type            & DDR4\_2400\_16x4                          \\ \hline
Size            & 32 GB                                       \\ \hline

Memorycpy latency & 1366.25 ns \\ \hline
LISA latency & 260.5 ns \\ \hline 
Shared-PIM latency & 158.25 ns \\ \hline 

\end{tabular}

\end{table}

\begin{figure}[t]

  \centering
 \includegraphics[scale=0.35]{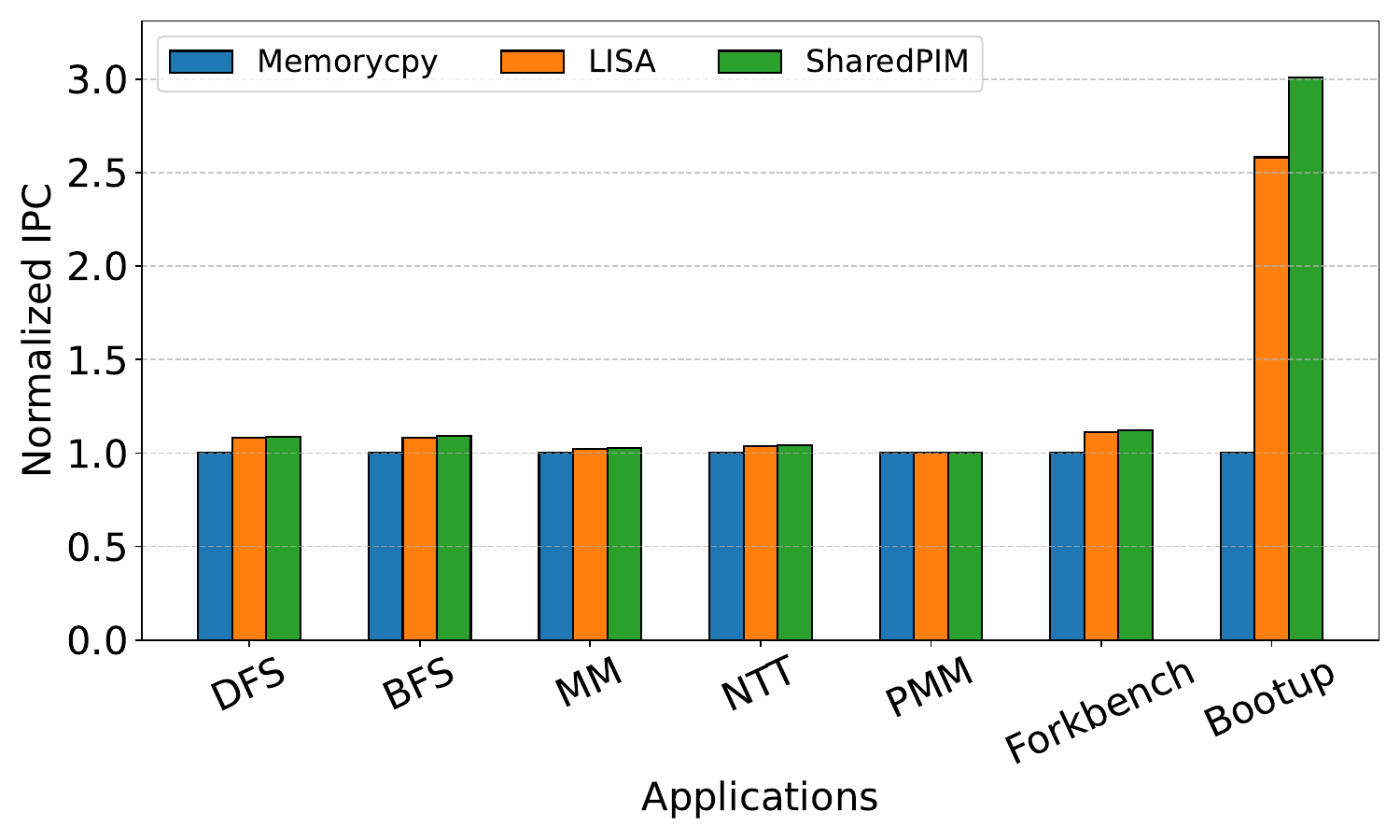}

  \caption{Normalized IPC comparison across various applications simulated in GEM5, with different memory transfer technologies. All results are normalized to memorycpy (set to 1.0) to highlight relative performance improvements.
  }
  \label{fig:non_pim}
% \vspace*{-5mm}
\end{figure}

To evaluate the benefits of Shared-PIM in supporting fast inter-subarray communication in regular DRAMs, and justify its overhead in non-PIM scenarios, we implemented Shared-PIM and LISA in the Gem5 system level simulator \cite{gem5} and used the SE model to simulate the applications tested in the previous section. The simulation setup is detailed in Table \ref{tab:Sim_para}. The results, shown in Fig.\ref{fig:non_pim}, present the normalized instructions per cycle (IPC) for each application benchmark evaluated in Sec. \ref{sec:system_level_eval}, with memorycpy performance set to 1.0 as a baseline for each application.

We also evaluated SPEC2006 \cite{spec2006}, Forkbench, and Bootup in Gem5 under non-PIM scenarios. We reduced the data size of the original SPEC2006 benchmark to prevent it from running indefinitely, making it feasible to simulate within a reasonable time frame. Forkbench creates 5000 child processes using fork(), performs floating-point calculations, and measures fork latencies. Bootup allocates 64MB of memory, performs computations and file I/O to measure system initialization and memory performance. Shared-PIM shows the highest benefit in Bootup due to its heavy memory transfers.

This evaluation demonstrates that Shared-PIM does not introduce any negative performance impact in non-PIM cases. While the primary benefit of Shared-PIM lies in its ability to enable parallel computation and data transfer in PIM architectures, the results confirm that it remains competitive in non-PIM scenarios without degrading performance.

\section{Conclusion}
%Placeholder

This paper introduced Shared-PIM, a solution that strategically leverages shared rows and sense amplifiers in in-DRAM PIM banks for synchronized PIM operations. Shared-PIM demonstrates a remarkable reduction in copy latency and energy consumption -- 5$\times$ and 1.2$\times$ lower than LISA \cite{lisa}, respectively. Additionally, it shows a notable improvement in basic computational tasks, with 1.4$\times$ faster addition and multiplication. These enhancements translate to substantial performance gains in real-world applications, as evidenced by the latency improvements of MM tasks by 40\%, PMM by 44\%, and NTT by 31\%. Moreover, for graph processing tasks like BFS and DFS, Shared-PIM achieves a 29\% improvement in speed with an area overhead of just 7.16\% compared to a pLUTo baseline.

\section*{Acknowledgment}
This work was supported in part by SUPREME, one of seven centers in JUMP 2.0, a Semiconductor Research Corporation (SRC) program sponsored by DARPA.


\ifCLASSOPTIONcaptionsoff
  \newpage
\fi



%\bibliographystyle{IEEEtran}


% \end{thebibliography}

\bibliographystyle{IEEEtran}
\bibliography{Bibliography}


% \end{IEEEbiography}


%\newpage



%\vfill


\end{document}